# Relativistic Drag and Emission Radiation Pressures in an Isotropic Photonic Gas

Jeffrey S. Lee[1]
Gerald B. Cleaver[2]

[1,2]Early Universe Cosmology and Strings Group, a Division of the Center for Astrophysics, Space Physics, and Engineering Research
[2]Department of Physics
Baylor University
One Bear Place
Waco, TX 76706

Jeff_Lee@Baylor.edu
Gerald_Cleaver@Baylor.edu

## Abstract

By invoking the relativistic spectral radiance, as derived by Lee and Cleaver [1], the drag radiation pressure of a relativistic planar surface moving through an isotropic radiation field, with which it is in thermal equilibrium, is determined in inertial and non-inertial frames. The forward- and rearward-directed emission radiation pressures are also derived and compared. A fleeting (inertial frames) or ongoing (some non-inertial frames) Carnot cycle is shown to exist as a result of an intra-surfaces temperature gradient. The drag radiation pressure on an object with an arbitrary frontal geometry is also described.

## 1. Introduction

This paper challenges the results of Balasanyan and Mkrtchian [2], in which the blackbody radiation drag on a relativistically moving mirror is calculated. To determine the relativistic photon drag, temperature inflation, Doppler shifting, and relativistic beaming of radiation in the direction of motion must all be considered.

Here, a $z$-axis-directed relativistic planar surface with an arbitrary absorptivity/emissivity, which is in thermal equilibrium with an isotropic thermalized photonic gas, is examined. Relativistic temperature transformations are accomplished by means of *inverse temperature*, a van Kampen-Israel future-directed timelike 4-vector. The body experiences radiation drag due to the momentum transfer from Doppler shifted and relativistically beamed photons incident upon its forward planar surface.

With increasing speed, the radiation field becomes progressively more anisotropic in the inertial frame, thus, creating a rising temperature difference between the forward and rearward thermally isolated surfaces of the body; this results in an emission radiation pressure gradient. If thermal sequestration of the forward and rearward surfaces is removed, the resulting temperature gradient across the body induces a short-lived Carnot cycle with a speed-dependent efficiency.

Additionally, the non-inertial reference frame case for drag and emission radiation pressures is established by considering the 4-acceleration. For surfaces which are not thermally isolated, a positive forward-to-rearward temperature gradient induces an "unending" (as long as there is acceleration) Carnot cycle because the surface temperatures never equalize. If the forward-to-rearward temperature gradient is negative (i.e. the body's speed is decreasing), the Carnot cycle ceases upon thermal equilibrium of the surfaces. Also, the drag radiation pressure acting upon a body with an arbitrary frontal surface geometry is determined.

## 2. The Drag Radiation Pressure

There are significant unresolved issues in the literature regarding temperature in relativistic thermodynamics [3], [4], [5], [6], [7], [8], [9]. Throughout the twentieth century, three published Lorentz group transformations have emerged: Temperature Deflation [5], [6] and Temperature Inflation [10], [11], [12] (which can be operationally quantified with a relativistic Carnot cycle [13], [14], [15]), and Temperature Invariance [16], [17], [18], [19].

The *empirical temperature* (a Lorentz invariant), which follows from the Zeroth Law of Thermodynamics, is a relativistic scalar that reflects the radiation rest frame and the observer frame as being in thermal equilibrium [20]. The validity of the Zeroth Law is necessarily independent of any reference to any thermodynamic property (including energy and entropy) [21].

The Second Law of Thermodynamics gives rise to the *absolute temperature*, which contains no angular dependence, and is the product of the radiation rest frame absolute temperature and the Lorentz factor.

Temperature transformations can be successfully realized by treating *inverse temperature* as a van Kampen-Israel future-directed timelike 4-vector. Although Przanowski and Tosiek [22][i], and Lee and Cleaver [3][ii] have demonstrated temperature inflation without making use of inverse temperature, a relativistic blackbody spectrum necessitates a consideration of angular dependence [1].

Balasanyan and Mkrtchian [2] calculate the drag pressure from blackbody radiation on a relativistic mirror. However, temperature, rather than the thermodynamically relevant inverse temperature, is transformed as a scalar. Also relativistic beaming and Doppler shifting of incident radiation, which are essential to describing the relativistic spectral radiance and energy density of a blackbody, are not considered.

### 2.1 Inertial Frames

In terms of the incident radiation pressure $P_{\text{INC}}$ [iii], the total radiation pressure $P_{\text{TOT}}$ on an opaque object in an inertial frame whose forward and rearward surfaces have an equal frequency-independent absorption/emission coefficient $\varepsilon$, is the sum of the radiation pressures due to absorption and reflection:

$$P_{\text{TOT}} = \varepsilon P_{\text{INC}} + 2(1-\varepsilon)P_{\text{INC}} = (2-\varepsilon)P_{\text{INC}} \tag{1}$$

Also,

[i] With a superfluidity gedanken experiment.
[ii] Rewriting the stress energy tensor using occupation number.
[iii] This is the total radiation pressure absorbed by a body with an absorptivity coefficient of 1.

$$P_{\mathrm{TOT}} = \frac{2-\varepsilon}{c}\int_0^\infty\int_{2\pi} B_\nu \cos^2\theta d\Omega d\nu \quad \text{iv} \tag{2}$$

where $c$ is the speed of light, $\theta$ is the angle of incoming radiation with respect to the direction vector of the object's motion, $v$ is the frequency of incoming photons, and $B_\nu$ is the spectral radiance in frequency space.

The solid angle integration is performed over 2π sr $(0 \le \phi \le \pi)$ because the photon drag acts only on the forward surface of the object. Relativistically, eq. (2) transforms to:

$$P'_{\mathrm{TOT}} = \frac{2-\varepsilon}{c}\int_0^\infty\int_{2\pi} B'_\nu \cos^2\theta' d\Omega' d\nu' \tag{3}$$

The relativistic spectral radiance in frequency space is given by [1]:

$$B'_\nu = \frac{\left(\dfrac{2h\nu^3}{c^2}\right)}{\exp\left[\dfrac{h\nu}{k_B}(\beta_t - \beta_z \cos\theta)\right] - 1}\left[\gamma(1 - V\cos\theta)\right]^{-3} \tag{4}$$

where:

$$\beta_t = \frac{1}{T_o\sqrt{1-V^2}} \tag{5}$$

$$\beta_z = \frac{V}{T_o\sqrt{1-V^2}} \tag{6}$$

$T_o$ is the proper temperature, $\beta_\mu = \dfrac{u_\mu}{T_o} = \beta_t - \beta_z \cos\theta$ is the van Kampen-Israel inverse temperature 4-vector, $u_\mu$ is the relative 4-velocity between the radiation and the observer. $\gamma = (1-V^2)^{-\frac{1}{2}}$ is the Lorentz factor, $V = \dfrac{u}{c}$ (fraction of light speed), and $h$ and $k_B$ are Planck's and Boltzmann's constants respectively.

Combining eqs. (3) and (4) yields[v]:

[iv] The $\cos^2\theta$ term results from the incoming radiation being incident in two dimensions.

$$P'_{\text{TOT}} = \frac{2-\varepsilon}{c}\int_0^{\infty}\int_0^{\frac{\pi}{2}} \frac{\left(\frac{2h\nu^3}{c^2}\right)}{\exp\left[\frac{h\nu}{k_B}\left(\beta_t - \beta_z\cos\theta\right)\right]-1}\left[\gamma\left(1 - V\cos\theta\right)\right]^{-3}\cos^2\theta\sin\theta\int_0^{\pi} d\phi d\nu \tag{7}$$

Since motion is along the $z$-axis, and the radiation is [azimuthally] isotropic, the relativistic spectral radiance can be separated from the azimuthal integral (i.e. $\beta_x = \beta_y = 0$).

By setting $\alpha = \frac{2h(2-\varepsilon)}{c^3}\left[\gamma\left(1 - V\cos\theta\right)\right]^{-3}\cos^2\theta\sin\theta\int_0^{\pi} d\phi = \frac{2\pi h(2-\varepsilon)}{c^3}\left[\gamma\left(1 - V\cos\theta\right)\right]^{-3}\cos^2\theta\sin\theta$, letting $R = \beta_t - \beta_z\cos\theta$, and setting $x = \frac{h}{k_B}R\nu$, $P'_{\text{TOT}}$ becomes:

$$P'_{\text{TOT}} = \int_0^{\infty}\int_0^{\frac{\pi}{2}} \frac{\alpha\nu^3}{e^x - 1} d\theta d\nu \tag{8}$$

The integral over frequency yields $\frac{\pi^4}{15}$.

Thus:

$$P'_{\text{TOT}} = \frac{\pi^4 k_B^4}{15h^4}\int_0^{\frac{\pi}{2}} \frac{\alpha}{R^4} d\theta \tag{9}$$

From eqs. (5), (6), and evaluating eq. (9):

$$P'_{\text{TOT}} = \frac{\pi^5(2-\varepsilon)k_B^4}{450c^3h^3}\left[\left(V^3 - 6V^2 + 15V - 20\right)\sqrt{1-V^2}\left(\frac{V+1}{V-1}\right)^3\right]T_o^4 \tag{10}$$

Eq. (10) is the total radiation pressure experienced by a relativistic body with arbitrary absorptivity. Expectedly, when $V \rightarrow 1$, relativistic beaming and temperature inflation cause $P'_{\text{TOT}} \rightarrow \infty$. Also, as expected, when $V = 0$:

$$P'_{\text{TOT}} = \frac{2\pi^5(2-\varepsilon)k_B^4 T_o^4}{45c^3h^3} = (2-\varepsilon)\frac{\sigma T_o^4}{3c} = (2-\varepsilon)\frac{S}{3c} = (2-\varepsilon)\frac{U}{3} \tag{11}$$

---

[v] The sin$\theta$ term emerges from the solid angle integration.

where $S$ is the radiation frame intensity, $\sigma = \dfrac{2\pi^5 k_B^4}{15h^3c^2}$ is the Stefan-Boltzmann constant, and $U$ is the energy density.

In the cases of a perfect blackbody $(\varepsilon = 1)$ and a perfect reflector $(\varepsilon = 0)$, $P' = \dfrac{U}{3}$ and $P' = \dfrac{2U}{3}$ respectively, as required.

Balasanyan and Mkrtchian [2] contend that the drag force (and drag pressure) are zero when an object is at rest in the photon field. However, this is not correct, as there will always exist a radiation pressure on all objects, regardless of whether they are in the same frame as the radiation source. Of course, an observer can be at rest in the frame of the radiation *source*, but is never at rest in the frame of the radiation source's *photons*. In an isotropic radiation field, the *net* drag radiation pressure will be zero when an axially symmetric object is stationary. However, this requires angular dependent Doppler shifting of the incident photons, which was not considered in [2]. This is shown in Figure 1.

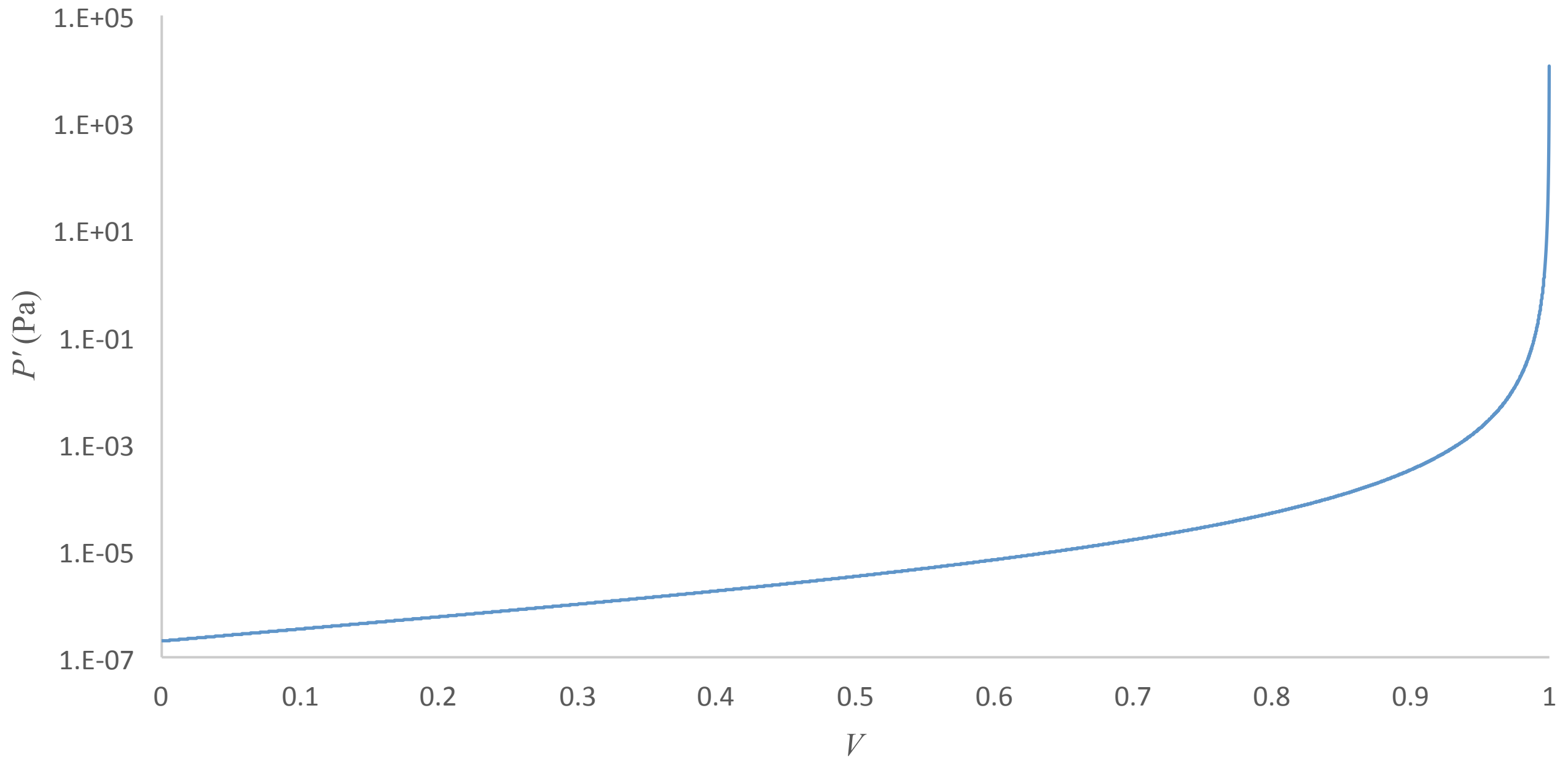

Figure 1: Drag radiation pressure versus speed for a perfectly absorbent blackbody ($\varepsilon = 1$) at 200 K. The $P'$-axis intercept represents the radiation pressure in the radiation frame and is 0.2006 $\mu$Pa.

### 2.2 Non-Inertial Frames

The radiation pressure in a non-inertial reference frame is determined with the 4-acceleration $a_\mu$, which is the proper time ($\tau$) derivative (denoted with dot notation) of the 4-velocity $(a_\mu = \dot{u}_\mu)$. Combining eqs. (12) and (10), and defining $A = \frac{a}{c}$ (the zeroth term of the acceleration 4-vector), the relativistic spectral radiance in a non-inertial frame is given by eq. (13).

$$V = \tanh(A\tau) \tag{12}$$

$$P'_{\text{TOT}} = \frac{\pi^5 (2-\varepsilon) k_B^4}{450 c^3 h^3} \left[ \frac{\left(\tanh^3(A\tau) - 6\tanh^2(A\tau) + 15\tanh(A\tau) - 20\right)}{\cosh(A\tau)} \left( \frac{\tanh(A\tau)+1}{\tanh(A\tau)-1} \right)^3 \right] T_o^4 \tag{13}$$

As $A\tau \to \infty$, $P'_{\text{TOT}} \to \infty$, and when $A\tau = 0$, the rest frame radiation pressure, given by (11), is recovered. A plot of eq. (13) is shown in Figure 2.

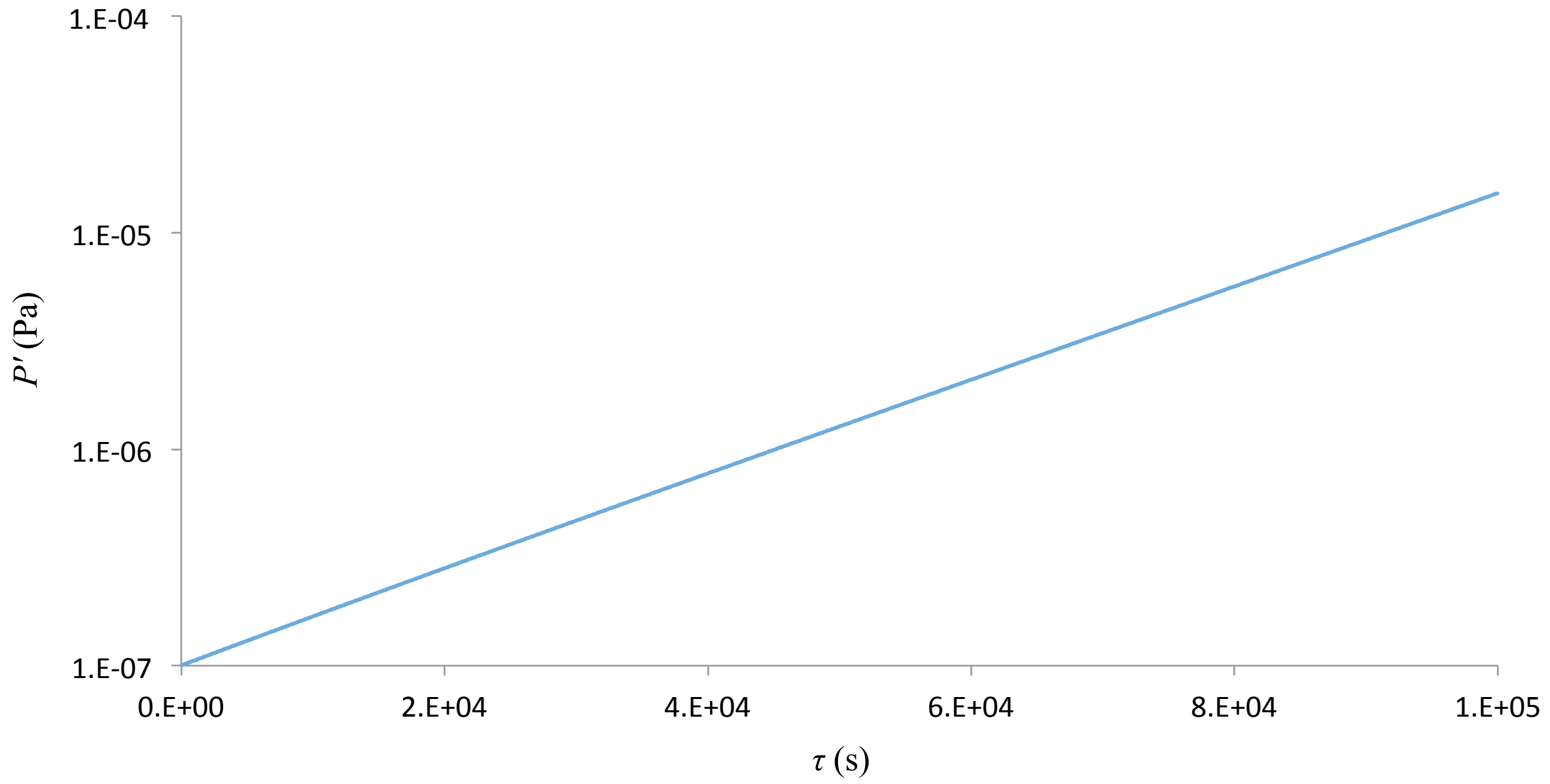

Figure 2: Radiation pressure as a function of proper time for a 200 K blackbody ($\varepsilon$ = 1). $A = 10^{-5}$. The $P'$-axis intercept represents the radiation pressure in the radiation frame and is 0.2006 $\mu$Pa.

In general, $V$ and $\theta$ will be time-dependent functions, and the time rate of change of radiation pressure is described by the proper time ($\tau$) derivative (denoted with dot notation) of the relativistic radiation pressure. Thus, $\dot{P} \equiv \frac{dP}{d\tau}$.

When eq. (13) is differentiated, the result is eq. (14), which is plotted in Figure 3.

$$\dot{P}'_{\mathrm{TOT}} = \frac{\pi^5 (2-\varepsilon) k_B^4}{450 c^3 h^3} A \left[ \frac{(\tanh(A\tau)+1)^2 (4\tanh^5(A\tau) - 20\tanh^4(A\tau) + 39\tanh^3(A\tau) - 35\tanh^2(A\tau) + 7\tanh(A\tau) + 105)}{\cosh(A\tau)(1-\tanh(A\tau))^3} \right] T_o^4 \quad (14)$$

Although perhaps unexpected, it is reasonable that $\dot{P}'_{\mathrm{TOT}}$ does not have angular dependence, even though $V$ and $\theta$ are time-dependent functions, because the strict definition of $\dot{P} \equiv \frac{dP}{d\tau}$ requires the proper time differentiation of a function from which all angular dependence has been removed by the $d\Omega$ integration. Defining $\dot{P}'_{\mathrm{TOT}} = \frac{2-\varepsilon}{c} \int_0^\infty \int_{2\pi} \dot{B}'_\nu \cos^2\theta' d\Omega' d\nu'$ does not adhere to the required proper time derivative definition of $\dot{P}'_{\mathrm{TOT}}$, in part because the order of the differentiation and integration cannot be inverted [1].

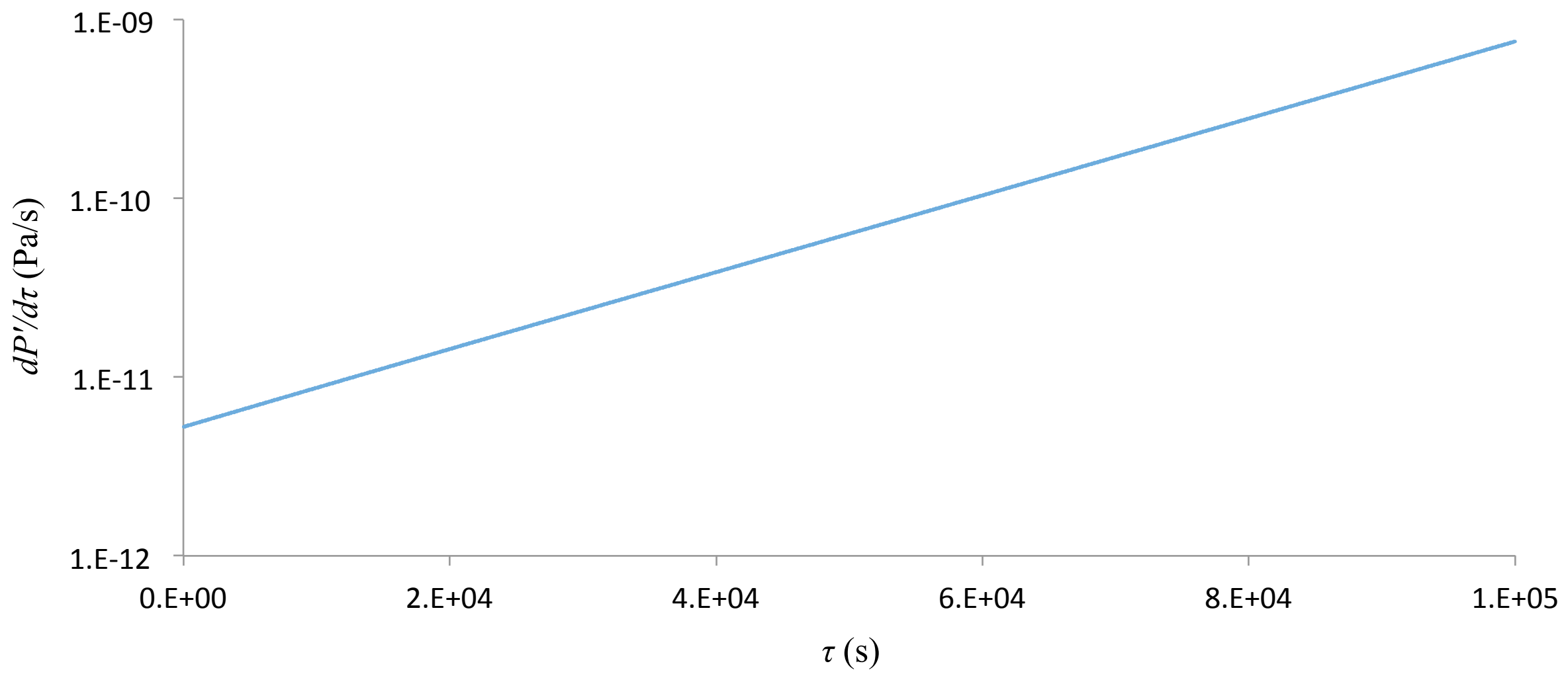

Figure 3: Time rate of change of radiation pressure as a function of proper time for a 200 K blackbody ($\varepsilon$ = 1). $A = 10^{-5}$. The $dP'/d\tau$-axis intercept is the initial instantaneous rate of change of the radiation pressure (5.27 pPa/s).

Also,

$$\dot{P}'_{\mathrm{TOT}}(\tau = 0) = \frac{7\pi^5(2-\varepsilon)k_B^4}{30c^3h^3}AT_o^4 = \frac{7\sigma}{4c}T_o^4 A = \frac{7}{4}UA \tag{15}$$

Expectedly, when $A = 0$, $\dot{P}'_{\mathrm{TOT}} = 0$.

## 3. Emission Radiation Pressure in Inertial Reference Frames

For a relativistic body in thermal equilibrium with an incident blackbody spectrum, the radiation pressure due to emission can be calculated. Since temperature is not a Lorentz invariant, the temperature at which the body equilibrates with the surrounding radiation is relativistically transformed. For bodies with forward and rearward surfaces which are in thermal isolation, the temperature gradient through the body is maintained, and there is a net forward-directed radiation pressure. If the surfaces are not thermally isolated, the temperatures equalize by means of internal heat conduction, and the net emitted radiation pressure approaches zero.

The emitted radiation pressure experienced by the body in its frame is:

$$P_{\mathrm{emission}} = \frac{S}{c} = \frac{\varepsilon\sigma}{cT_o^{-4}} \tag{16}$$

### 3.1 Radiation Pressure due to Forward Emitted Radiation

The pressure due to radiation emitted in the direction of motion $(\theta = 0)$ $P^{\rightarrow}_{\mathrm{emission}}$, can be determined from eqs. (5) and (6), and by replacing $T_o^{-1}$ with the inverse temperature 4-vector $\beta_\mu = \frac{u_\mu}{T_o} = \beta_t - \beta_z \cos\theta$:

$$P^{\rightarrow}_{\mathrm{emission}} = \frac{\varepsilon\sigma}{c}\left(\frac{1+V}{1-V}\right)^2 T_o^4 \tag{17}$$

Clearly, when $V = 0$, the rest frame Stefan-Boltzmann Law (divided by *c*) is recovered. As $V \rightarrow 1$, the emission radiation pressure becomes infinite.

For $V \sim 0$, a series expansion of eq. (17) yields:

$$P^{\rightarrow}_{\mathrm{emission}} \sim \frac{\varepsilon\sigma}{c}\left(1 + 4V + 8V^2 + O\left(V^3\right)\right)T_o^4 \tag{18}$$

The emissivity which produces equal drag and forward-directed emission radiation pressures is independent of temperature, can easily be calculated by equating eqs. (10) and (17), and is given by:

$$\varepsilon = \frac{2\left(V^3 - 6V^2 + 15V - 20\right)}{\left(\frac{-60}{\sqrt{1-V^2}}\right) + V^3 - 6V^2 + 15V - 20} \tag{19}$$

A plot of eq. (19) is shown in Figure 4.

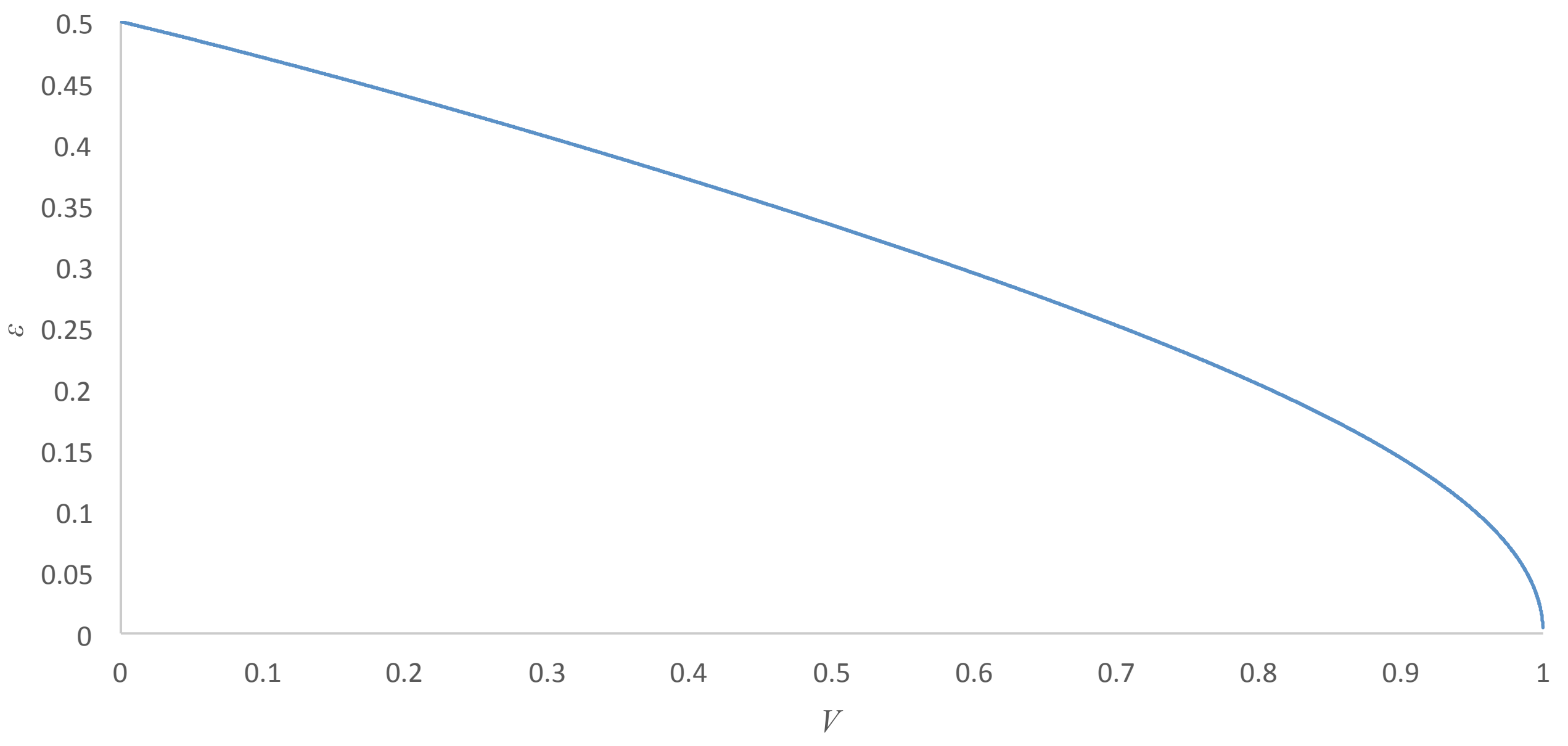

Figure 4: Emissivity versus speed for equal drag and forwarded-directed emission radiation pressures.

The emissivities which produce equal radiation pressures when $V = 0$ and $V \to 1$ are not surprising 0.5 and 0 respectively. The speed which produces equal drag and emission radiation pressures for a given emissivity is also independent of temperature, and is given by the non-trivial solution to:

$$\left(V^3 - 6V^2 + 15V - 20\right)\sqrt{1-V^2} = \frac{60\varepsilon}{\varepsilon - 2} \tag{20}$$

### 3.2 Radiation Pressure due to Rearward Emitted Radiation

The pressure due to radiation emitted in the direction opposite of motion $\left(\theta = 180^o\right)$ $P^{\leftarrow}_{\text{emission}}$, can be determined again from eqs. (5) and (6), and by replacing $T_o^{-1}$ with the inverse temperature 4-vector $\beta_\mu = \frac{u_\mu}{T_o} = \beta_t - \beta_z \cos\theta$:

$$P^{\leftarrow}_{\text{emission}} = \frac{\varepsilon\sigma}{c}\left(\frac{1-V}{1+V}\right)^2 T_o^4 \tag{21}$$

As is the case with eq. (17), when $V = 0$, the rest frame Stefan-Boltzmann Law is recovered. As $V \to 1$, the emission radiation pressure becomes zero.

For $V \sim 0$, a series expansion of eq. (21) gives:

$$P^{\leftarrow}_{\text{emission}} \sim \frac{\varepsilon\sigma}{c}\left(1 - 4V + 8V^2 + O\left(V^3\right)\right)T_o^4 \tag{22}$$

A comparison of drag radiation pressure and emission radiation pressures is shown in Figure 5.

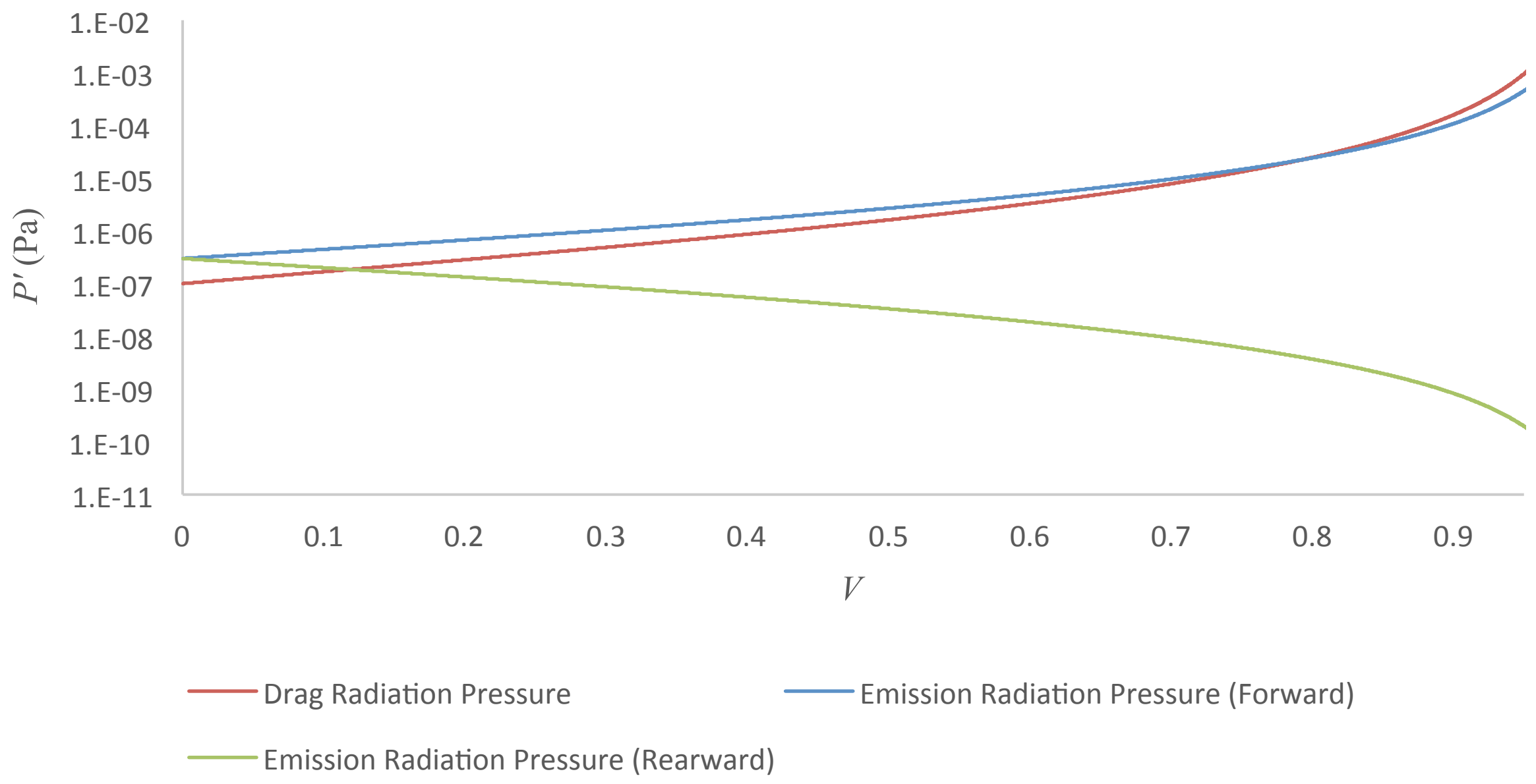

Figure 5: Drag, forward-directed, and rearward-directed emission radiation pressures for a 200 K blackbody ($\varepsilon = 1$). The drag and forward-directed radiation pressures are equal for a speed of ~0.7927*c*. The drag and rearward-directed radiation pressures are equal for a speed of ~0.1189*c*. Clearly, the forward-directed and rearward-directed radiation pressures are equal only when the object is stationary.

This result is in contradiction to the calculation of Balasanyan and Mkrtchian [2], in which a speed of ~0.1$c$ results in equal drag and emission radiation pressures. However, they show that as $V \to 1$, the drag radiation pressure vastly exceeds the emission radiation pressure, with which Figure 5 is in agreement.

Although the forward-directed emission radiation pressure is markedly greater than the drag radiation pressure for the expected speeds of a solar sail spacecraft $\left(V \sim 10^{-4} - 10^{-3}\right)$, the difference in incident flux, resulting from incident radiation anisotropy, results in a greater rearward-directed thrust from both forward-directed emission radiation pressure and the rearward-directed drag radiation pressure.

The emissivity which produces equal forward drag radiation pressure and rearward emission radiation pressure is also independent of temperature, can easily be calculated by equating eqs. (10) and (23), and is given by:

$$\varepsilon = \frac{2\left(V^3 - 6V^2 + 15V - 20\right)\sqrt{1-V^2}\left(\frac{V+1}{V-1}\right)^5}{60 + \left(V^3 - 6V^2 + 15V - 20\right)\sqrt{1-V^2}\left(\frac{V+1}{V-1}\right)^5} \tag{23}$$

A plot of (23) is shown in Figure 6.

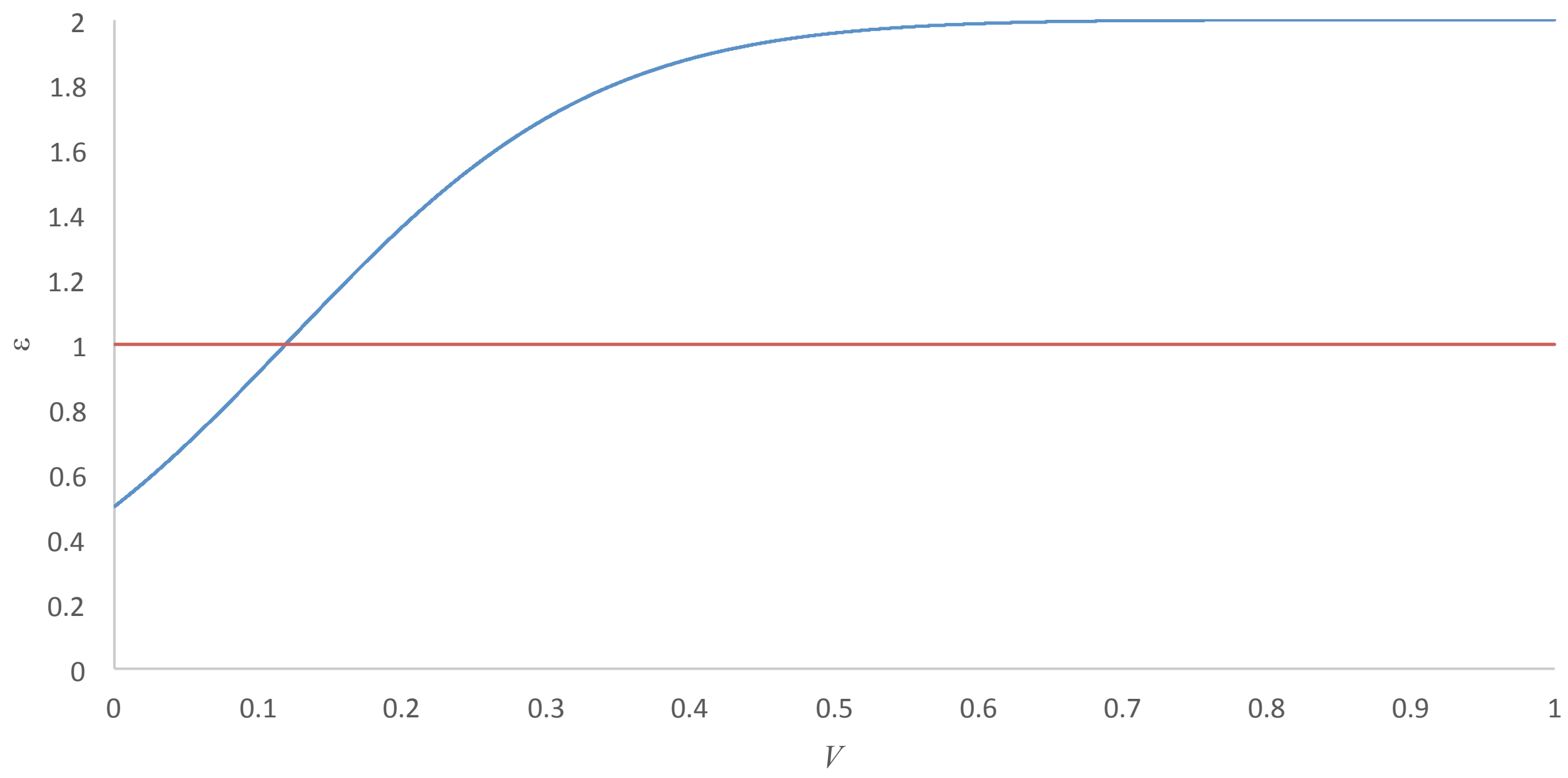

Figure 6: Emissivity versus speed for equal drag and rearward-directed emission radiation pressures. The region of the graph above the red line represents non-physical emissivities, and thus speeds for which it is not possible for these radiation pressures to be equal. As expected from Figure 5, the cutoff speed at which the radiation pressures are equal is ~0.1189$c$. The emissivity which produces equal radiation pressures when $V = 0$ is, not surprisingly, 0.5.

### 3.3 Emitted Radiation Pressure in Non-Inertial Frames

The forward- and rearward-directed emission radiation pressures in a non-inertial frame can be determined by combining eq. (12) with eqs. (17) and (21) respectively.

Therefore,

$$P^{\overrightarrow{\leftarrow}}_{\text{emission}} = \frac{\varepsilon\sigma}{c}\left(\frac{1 \pm \tanh(A\tau)}{1 \mp \tanh(A\tau)}\right)^2 T_o^4 \tag{24}$$

A plot of eq. (24) is shown in Figure 7.

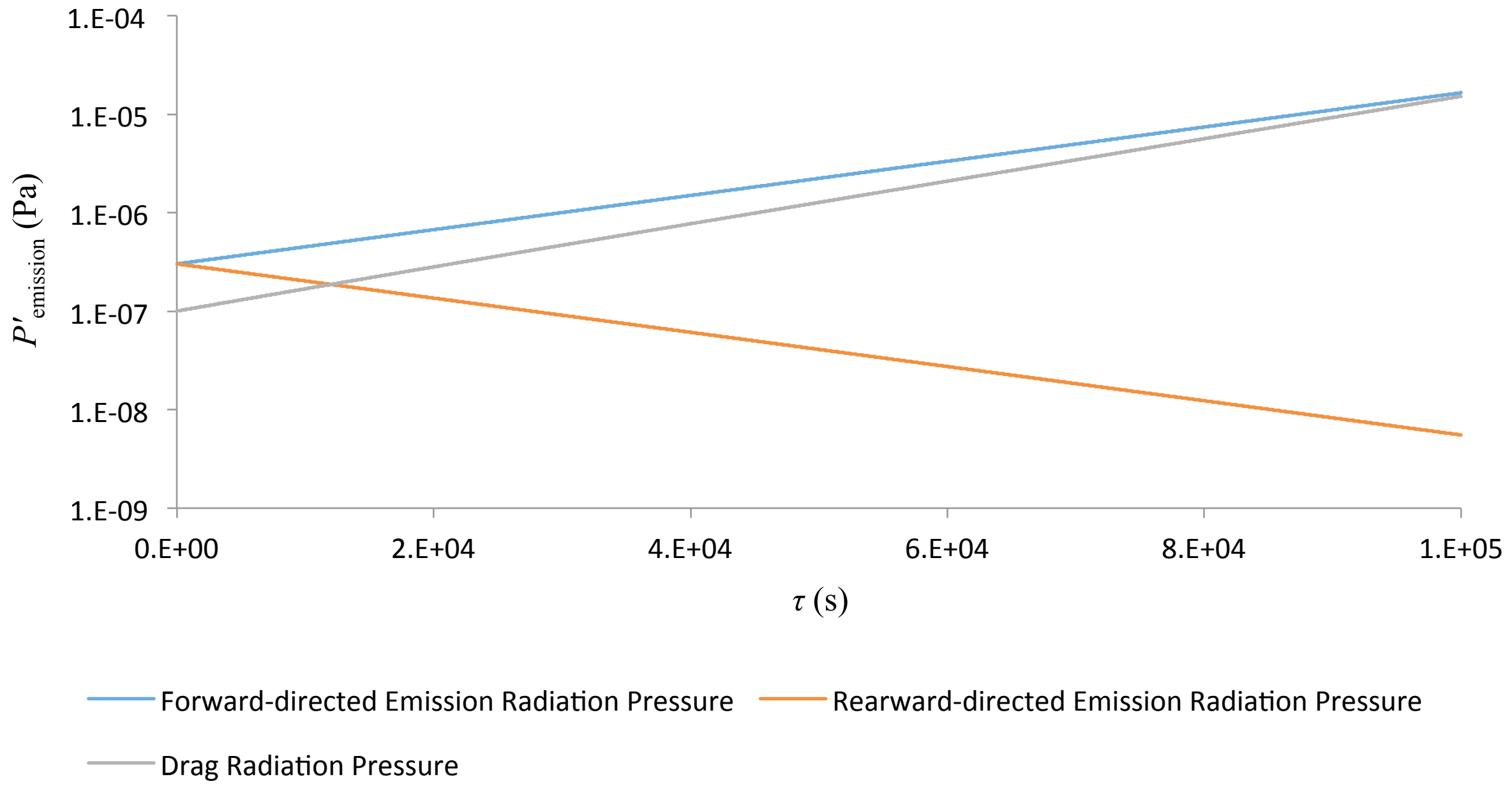

Figure 7: Drag, forward-directed, and rearward-directed emission radiation pressures for a 200 K blackbody ($\varepsilon = 1$). $A = 10^{-5}$. The drag and forward-directed radiation pressures are equal at a proper time of $\sim 1.0835\times 10^5$ s. The drag and rearward-directed radiation pressures are equal for a time of $\sim 1.2001\times 10^4$ s. Clearly, the forward-directed and rearward-directed radiation pressures are equal only when the blackbody is stationary.

The time rate of change of the forward- and rearward-directed emission radiation pressures can easily be determined.

$$\dot{P}_{\text{emission}} \equiv \frac{dP_{\text{emission}}}{d\tau} \tag{25}$$

Therefore, from eq. (24):

$$\dot{P}^{\rightarrow}_{\underset{\text{emission}}{\leftarrow}} = \pm\frac{4\varepsilon\sigma}{c}A\left(\frac{1\pm\tanh(A\tau)}{1\mp\tanh(A\tau)}\right)^2 T_o^4 = \pm 4AP^{\rightarrow}_{\underset{\text{emission}}{\leftarrow}} \tag{26}$$

Clearly, when $A\tau = 0$, $\dot{P}^{\rightarrow}_{\underset{\text{emission}}{\leftarrow}} = \pm\frac{4\varepsilon\sigma}{c}A$, and when $A\tau \to \infty$, $\dot{P}^{\rightarrow}_{\text{emission}} \to \infty$ and $\dot{P}^{\leftarrow}_{\text{emission}} \to 0$.

### 3.4 Temperature Gradient-Induced Carnot Cycle

The forward and rearward surfaces of the blackbody experience temperature inflation and deflation respectively. Thus, their respective absolute temperatures are $T_H$ $(\theta = 0)$ and $T_C$ $(\theta = 180^o)$. If the surfaces of the body are not thermally isolated, heat will flow from the forward to the rearward surface, and a Carnot cycle results.

The efficiency $\eta$ of a Carnot cycle in the inertial frame of the body is:

$$\eta = 1 - \frac{T_C}{T_H} \tag{27}$$

where $T_C$ and $T_H$ are the cold and hot reservoir temperatures respectively.

From eqs. (5) and (6), and by replacing $T_o^{-1}$ with the inverse temperature 4-vector $\beta_\mu = \frac{u_\mu}{T_o} = \beta_t - \beta_z\cos\theta$, $\eta$ in terms of $V$ can easily be determined.

$$\eta = \frac{2V}{1+V} \tag{28}$$

Expectedly, when $V = 0$, the efficiency is zero, and the Carnot cycle does not exist due to the zero temperature gradient. As $V \to 1$, $\eta \to 1$.

However, without thermal isolation in an inertial frame (or in a non-inertial frame with negative acceleration, see below), the surface temperatures eventually equilibrate, and the Carnot cycle shuts down in a proper time which is dependent on the thermal diffusivity of the body.

In a non-inertial frame, the Carnot efficiency is obtained by combining eqs. (12) and (28), is given by eq. (29) [vi], and is plotted in Figure 8.

$$\eta = \frac{2\tanh(|A|\tau)}{1+\tanh(|A|\tau)} \tag{29}$$

---

[vi] $|A|$ is required because $V = A\tau \geq 0$.

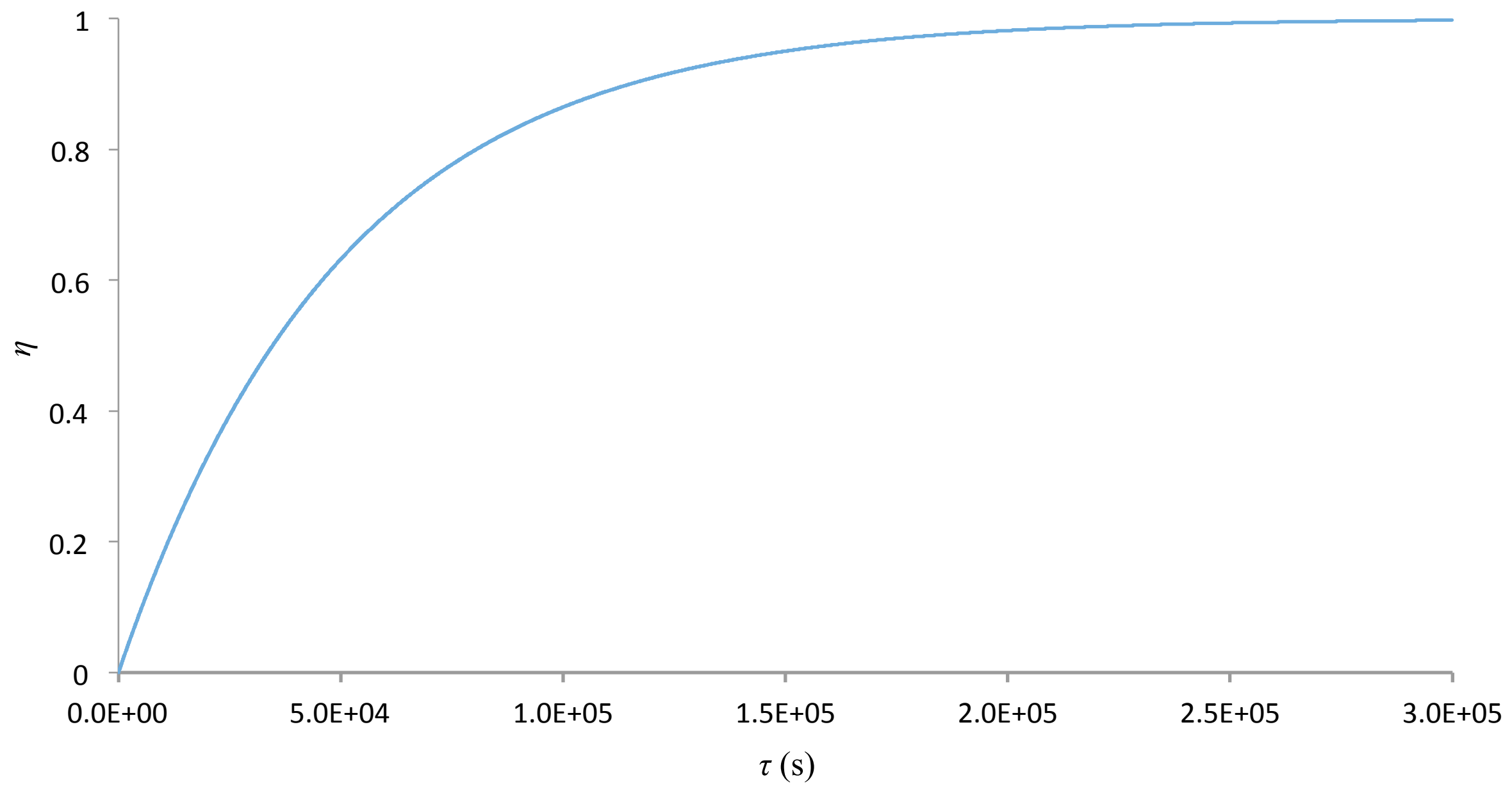

Figure 8: Efficiency of a Carnot cycle as a function of proper time for a blackbody with positive acceleration and a radiation frame temperature of 200 K. $A = 10^{-5}$.

In the frame of the body, the time rate of change of the Carnot cycle efficiency is the proper time derivative of eq. (29), and is plotted in Figure 9 for positive acceleration and in Figure 10 for negative acceleration.

$$\dot{\eta} = 2A\left(\frac{1-\tanh(A\tau)}{1+\tanh(A\tau)}\right) \tag{30}$$

With positive acceleration and consequently increasing speed, the Carnot cycle continues indefinitely because the constantly rising temperature of the forward surface and the finite time required for heat flow through the body collectively prevent the forward and rearward surfaces from ever achieving thermal equilibrium, as $\eta \to 1$ and $\dot{\eta} \to 0$.

However, in the case of negative acceleration, the Carnot efficiency is decreasing in time, as $\eta \to 0$ and $\dot{\eta} \to -\infty$. Therefore, the surfaces eventually reach thermal equilibrium, and the Carnot cycle consequently shuts down.

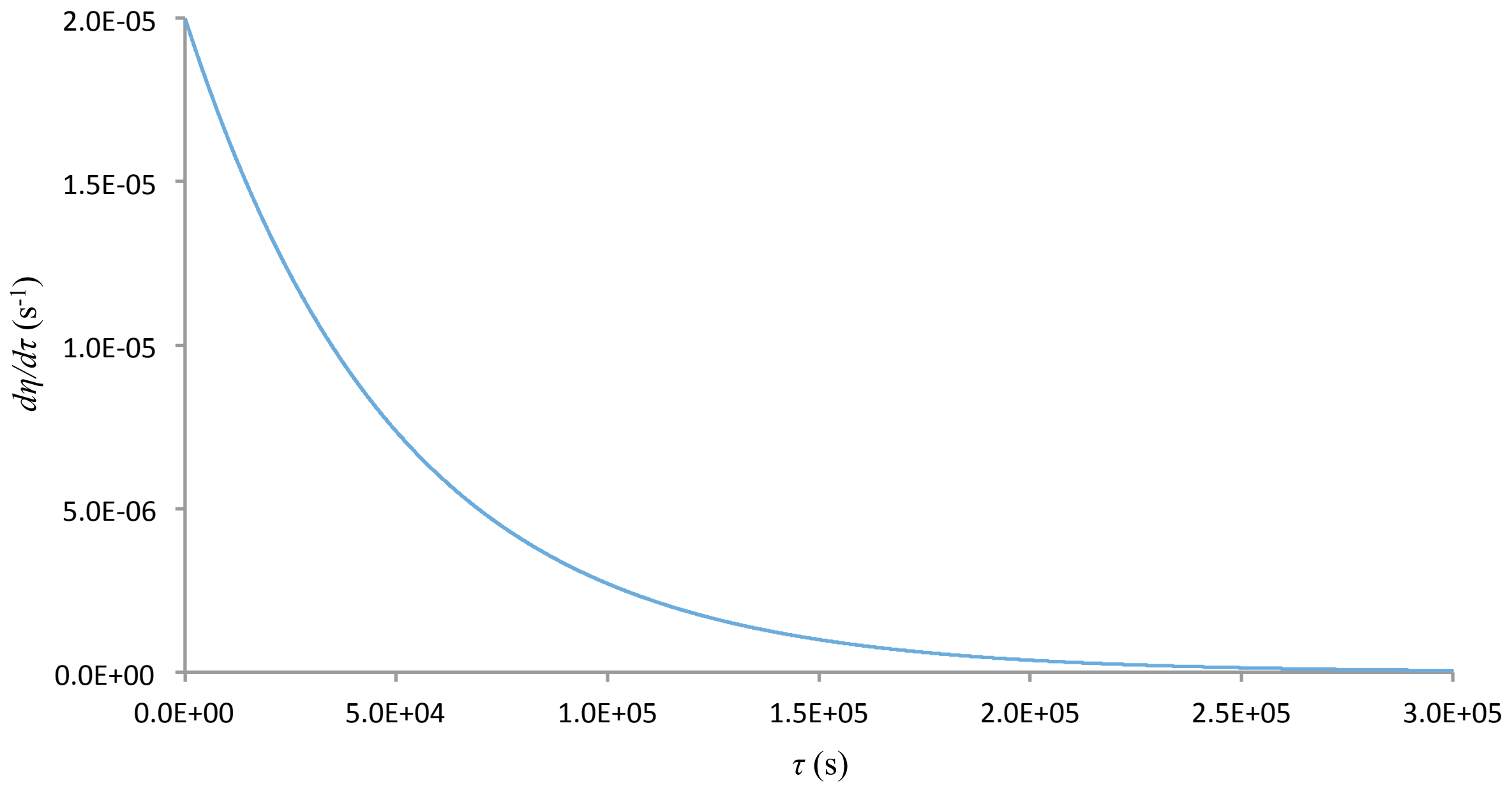

Figure 9: Plot of the time rate of change of the Carnot efficiency as a function of proper time for a 200 K blackbody. $A = 10^{-5}$.

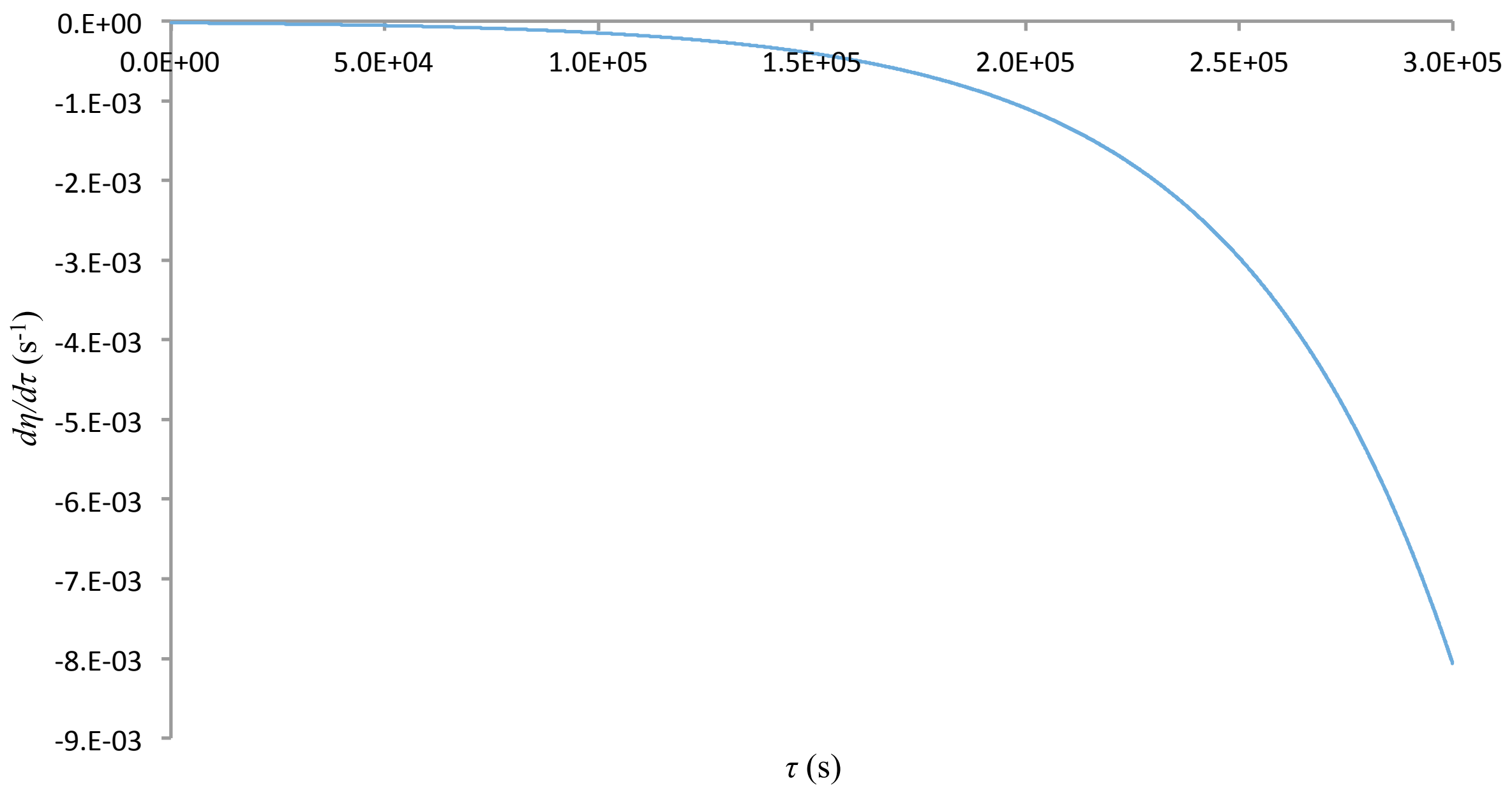

Figure 10: Plot of the time rate of change of the Carnot efficiency as a function of proper time for a 200 K blackbody. $A = -10^{-5}$.

## 4. Drag Radiation Pressure on a Surface of Arbitrary Geometry

For an object with an arbitrary frontal surface geometry moving relativistically in an inertial reference frame and in thermal equilibrium with an isotropic radiation field, the drag radiation pressure can be calculated.

If the forward surface is described by the function $F(x,y,z)=0$, then the normal $\hat{n}$ to the surface is clearly:

$$\hat{n} = \frac{\nabla F}{|\nabla F|} \tag{31}$$

The total drag radiation pressure $P'_{\text{TOT}}$ is merely the scalar product of the total drag radiation pressure vector $\vec{P}'_{\text{TOT}}$ and the surface normal vector.

$$P'_{\text{TOT}} = \vec{P}'_{\text{TOT}} \cdot \hat{n} \tag{32}$$

Since motion occurs only along the $z$-axis:

$$P'_{\text{TOT}} = \frac{\left|\vec{P}'_{\text{TOT}}\right| \hat{z} \cdot \nabla F}{|\nabla F|} \tag{33}$$

If the frontal surface geometry is a plane parallel to the $x$-$y$ plane, then $\nabla F = \hat{z}$, and $P'_{\text{TOT}}$ is given by eq. (10).

If the reference frame in non-inertial, then the time rate of change of the drag radiation pressure is:

$$\dot{P}'_{\text{TOT}} = \frac{d}{d\tau}\left[\frac{\left|\vec{P}'_{\text{TOT}}\right| \hat{z} \cdot \nabla F}{|\nabla F|}\right] \tag{34}$$

which yields,

$$\dot{P}'_{\text{TOT}} = \frac{\left|\dot{\vec{P}}'_{\text{TOT}}\right| \hat{z} \cdot \nabla F}{|\nabla F|} + \left|\vec{P}'_{\text{TOT}}\right| \hat{z} \cdot \left[\frac{\nabla \dot{F}}{|\nabla F|} - \frac{\nabla F}{|\nabla F|^3}\left(F_x \dot{F}_x + F_y \dot{F}_y + F_z \dot{F}_z\right)\right] \tag{35}$$

where $F_i$ are the partial derivatives of $F$, and $\left|\dot{\vec{P}}'_{\text{TOT}}\right|$ is given by:

$$\left|\dot{\vec{P}}'_{\text{TOT}}\right| = \left|\frac{\pi^5 (2-\varepsilon) k_B^4}{450 c^3 h^3} A \left[\frac{(\tanh(A\tau)+1)^2 \left(4\tanh^5(A\tau) - 20\tanh^4(A\tau) + 39\tanh^3(A\tau) - 35\tanh^2(A\tau) + 7\tanh(A\tau) + 105\right)}{\cosh(A\tau)(1-\tanh(A\tau))^3}\right] T_o^4\right| \tag{36}$$

which is just the magnitude of eq. (14). If the frontal surface geometry is a plane parallel to the *x*-*y* plane, then $\nabla F = \hat{z}$, and $\dot{P}'_{\mathrm{TOT}}$ is given by eq. (14).

## Conclusions

The relativistic radiation pressure due to drag, experienced by a thermally equilibrated planar surface moving through an isotropic radiation field in both inertial and non-inertial reference frames, has been determined. These have been compared to the forward- and rearward-directed emission radiation pressures caused by the thermally isolated surfaces of the body. Also, the speeds at which the emission and drag radiation pressures are equal were determined and shown to be dependent on the emissivity, and independent of the temperature of the photonic gas.

An inertial body (or a non-inertial body with a negative acceleration), through which thermal conduction can occur, will experience a Carnot cycle between its forward and rearward surfaces, which will continue until the surface temperatures equalize in a time which is dependent on the body's thermal diffusivity. However, for such a body in a non-inertial frame with a positive acceleration, there will be a continuous Carnot cycle with a steadily increasing acceleration-dependent efficiency.